\documentclass{aa}  

\usepackage{graphicx}
\usepackage{txfonts}
\usepackage{booktabs}
\usepackage{hyperref}
\usepackage{siunitx}
\usepackage{makecell}
\DeclareSIUnit\year{yr}
\DeclareSIUnit\erg{erg}
\DeclareSIUnit\draine{Draine}
\begin{document}

   \title{Chemulator: Fast, accurate thermochemistry for dynamical models through emulation}
  \author{J. Holdship \inst{1,2}
          \and
         S. Viti \inst{1,2}
        \and        
          T. J. Haworth \inst{3}
          \and
          J. D. Ilee \inst{4}
          }

  \institute{Leiden Observatory, Leiden University, PO Box 9513, 2300 RA Leiden, The Netherlands\\
              \email{holdship@strw.leidenuniv.nl}
         \and
             Department of Physics and Astronomy, University College London, Gower Street, WC1E 6BT, London, UK
        \and
            Astronomy Unit, School of Physics and Astronomy, Queen Mary University of London, London E1 4NS, UK
        \and
            School of Physics \& Astronomy, University of Leeds, Woodhouse Lane, Leeds, LS2 9JT, UK
             }


  \abstract
   {Chemical modelling serves two purposes in dynamical models: accounting for the effect of microphysics on the dynamics and providing observable signatures. Ideally, the former must be done as part of the hydrodynamic simulation but this comes with a prohibitive computational cost which leads to many simplifications being used in practice. }
   {To produce a statistical emulator that replicates a full chemical model capable of solving the temperature and abundances of a gas through time. This emulator should suffer only a minor loss of accuracy over including a full chemical solver in a dynamical model but would have a fraction of the computational cost.}
   {The gas-grain chemical code UCLCHEM was updated to include heating and cooling processes and a large dataset of model outputs from possible starting conditions was produced. A neural network was then trained to map directly from inputs to outputs}
  {Chemulator replicates the outputs of UCLCHEM with an overall mean squared error (MSE) of \num{1.7e-4} for a single time step of 1000 yr and is shown to be stable over 1000 iterations with an MSE of \num{3e-3} on the log scaled temperature after one time step and \num{6e-3} after 1000 time steps. Chemulator was found to be approximately 50,000 times faster than the time dependent model it emulates but can introduce a significant error to some models. }
   {}

   \maketitle
   
%

\section{Introduction}
Hydrodynamic simulations are a cornerstone of theoretical astrophysics. However pure hydrodynamics is often insufficient, for example in scenarios where radiation, magnetic fields and/or the chemical evolution plays an important role in setting the dynamics. Furthermore, the chemical composition of an astrophysical system\ is important for understanding observables.\par
Given the above there has been substantial, sustained effort to develop numerical methods that are capable of dealing with the combination of required spatial resolution, time step, and microphysics. However, even with a pure hydrodynamics calculation there are many scenarios where the required resolution for convergence alone makes simulations prohibitively expensive \citep[e.g. the now resolved problem of convergence in the behaviour of self-gravitating discs, see][]{2011MNRAS.411L...1M, 2012MNRAS.427.2022M, 2015MNRAS.451.3987Y, 2017ApJ...847...43D}.  When additional microphysics such as radiative transfer or chemistry are included, which are typically orders of magnitude more computationally expensive than hydrodynamics \textit{per time step} (and may also reduce the possible time step that can be taken) we find that most problems cannot be feasibly addressed. 

Nevertheless, approximations have been made that have permitted radiation/chemical/magnetic hydrodynamic applications on the scales of cosmology/galaxies \citep[e.g.][]{2009MNRAS.400.1283I, 2012MNRAS.424L..11F, 2018MNRAS.473.4077P} through ISM evolution and star formation \citep[e.g.][]{2012MNRAS.421..116G, 2018MNRAS.477.5422A, 2021arXiv210101193A}, protoplanetary disc evolution and planet formation \citep[e.g.][]{2017MNRAS.472..189I, Haworth2019TheDiscs, 2020ApJ...899..134K} right down to exoplanet atmospheres \citep[e.g.][]{2012AA...546A..43V, 2018ApJ...855L..31D}. However by necessity these approximations sacrifice accuracy to make the problem computationally feasible. Typical approximations regarding the chemistry are to assume equilibirum (which could be inaccurate both for the temperature/dynamics and composition/observables) and to use severely reduced networks with what are thought to be the key components for any given problem. However it is difficult to know a priori to what extent these approximations are reasonable for any given application. Furthermore, there may even be qualitatively new processes that are overlooked through simplified approaches to the microphysics \citep[e.g. this was discussed and reviewed in the context of protoplanetary discs by][]{2016PASA...33...53H}. 

One alternative to addressing the large computational cost is to employ an efficiency boosting tactic. 
For example, on the fly calculations can be used to determine reactions or species in a chemical network that are unimportant in any given timestep, allowing them to be left out of the calculation \citep{Grassi2012}. This can give considerable speed up in models where the overhead from the ``importance calculations'' is outweighed by the speed up of the chemical solver. Alternatively, large grids of models can be calculated in advance and a look up table interpolated over in dynamical models \citep[eg.][]{Ploeckinger2020RadiativeFields}. Another innovation has been to combine multiple reactions and particles into smaller sets of meta-reactions and meta-particles \citep[e.g.][]{1999ApJ...524..923N}. However, even with these innovations there are still many problems that remain out of reach. 

In spite of the above, it may be possible to obtain a huge decrease in computational time whilst maintaining high accuracy through an emulator. An emulator in this context is a statistical model which returns the same outputs as a numerical model for the same inputs but is much faster to evaluate. These have been used to speed up parameter inference from radiative transfer and chemical models \citep{deMijolla2019IncorporatingEmulation} as well as to solve the thermochemistry of a small reaction network \citep{Grassi2011}.\par

In this work, the code Chemulator\footnote{\url{https://github.com/uclchem/Chemulator}} is developed. The objective is to develop an emulator for solving the chemistry and gas temperature with sufficient accuracy and speed to render a host of new problems attainable. The emulator should take the state of the gas at a time  $t$ as an input and returns the state of the gas (composition, temperature) at a time $\Delta t$ later. This will result in a model with approximately the accuracy of a full chemical solver with the much smaller computational cost of evaluating a neural network.

To achieve this, a zero dimensional model that includes a detailed treatment of the relevant chemical and thermal processes was produced and is described in Section~\ref{sec:model}. This is then used to train the network, with the training process aided by a dimensionality reduction described in Section~\ref{sec:dimensionality}. The model training and basic performance metrics are presented in Section~\ref{sec:emulator} and the work is summarized in Section~\ref{sec:conclusion}.
\section{The Model}
\label{sec:model}
\subsection{Model Requirements}
The model should take the physical parameters and chemical abundances of a gas at time $t$ and return the temperature and abundances at a time $t+\Delta t$. It is intended to be employed as a subgrid model in 2/3D dynamical simulations. This results in a crucial requirement on the list of the model specifications due to two competing constraints.\par
Firstly, in order for the emulator to work, the model cannot rely on information about the object beyond the initial conditions of a given time step. For example, radiative transfer solvers typically iteratively solve the optical depths along "rays" from a given position. However, the emulator will not have access to internal values of the model such as those optical depths and so the model cannot rely on this method.\par
However, in order for the emulator to be effective at its intended task, non-local effects such as line cooling should approximate their effect in multi-dimension models. Therefore a set of inputs must be found that allow the 0D treatment to recover the temperatures of a non-local treatment. In the following sections, the solution to this is discussed and the model is benchmarked against a 1D model.\par
\subsection{Model Description}
The time dependent gas-grain chemical code UCLCHEM \citep{Holdship2017} was adapted to serve as the model to be emulated. It has been modified to include heating processes from UCL-PDR \citep{Bell2005,Priestley2017ModellingNebula}, molecular line cooling \citep{Bell2005,de1980hydrostatic} and other cooling processes from  KROME \citep{Grassi2014}.  See Tables~\ref{table:cooling} and \ref{table:heating} for a list of implemented heating and cooling processes. 
\begin{table}[]
    \centering
    \begin{tabular}{lr}
    \toprule
        \textbf{Cooling Process} & \textbf{Source}\\
        \midrule
        \multicolumn{2}{c}{\textbf{KROME}}\\
        H, He, He$^+$ collisional ionization & \citet{Cen1992}\\
        H$^+$,He$^+$,He$^{2+}$ recombination & \citet{Cen1992}\\
        He dielectric recombination & \citet{Cen1992} \\
        H collisional excitation & \citet{Cen1992}\\
        He collisional excitation & \citet{Cen1992}\\
        He+ collisional excitation & \citet{Cen1992}\\
        H2 collisional dissociation & \makecell[r]{\citet{Martin1998CollisioninducedDensities}\\ \citet{Glover2007SimulatingConditions}}\\
        Bremsstrahlung all ions & \citet{Cen1992}\\
        Compton cooling & \citet{Cen1992}\\
        Continuum & \citet{Hirano2013RadiativeFormation} \\
        Collisionally Induced Emission & \citet{Hirano2013RadiativeFormation}\\

        &\\
                \multicolumn{2}{c}{\textbf{UCL\_PDR}} \\
        LVG line cooling& \citet{de1980hydrostatic} \\
        \makecell[l]{H2 single pseudo level\\ vibrational cooling} & \citet{Rollig2006CIIRegions}\\
        Dust cooling (and heating) & \citet{Hollenbach1979MoleculeProcesses}\\
        Important endothermic Reactions & \citet{Bell2005} \\
        \bottomrule
    \end{tabular}
    \caption{Cooling Processes in the model. The processes are listed with their original source and grouped by the source for the treatment used in the model. }
    \label{table:cooling}
\end{table}
The model consistently solves the time-dependent gas chemistry and temperature for a single position in a cloud. The required inputs are the initial chemical abundances and the physical properties of the cloud at that position as listed in Table~\ref{table:inputs}. The outputs are the gas temperature and chemical abundances at a later time.\par
\begin{table}[]
    \centering
    \begin{tabular}{lr}
        \toprule
        \textbf{Heating Process} & \textbf{Source}\\
        \midrule
        Photoelectric Heating  & \citet{Weingartner2001} \\
        H2 Formation & \citet{Hollenbach1999} \\
        H2 Photodissociation & \citet{Hollenbach1979MoleculeProcesses}  \\
        H2 FUV pumping & \citet{Hollenbach1979MoleculeProcesses}\\
        C ionization & \citet{Kamp2001OnStars} \\
        CR heating & \citet{Goldsmith2001MolecularCores} \\
        Important exothermic Reactions & \citet{Bell2005} \\
        Gas-grain Collisions & \citet{Burke1983THETRAPPING}\\
        Turbulent decay & \citet{Black1987HeatingGas}  \\ 
        \bottomrule
    \end{tabular}
    \caption{Heating Processes in the model with original reference. The code and processes were taken from UCL\_PDR \citep{Bell2005}.}
    \label{table:heating}
\end{table}
The requirement that the model solves a single position independently of other positions has been satisfied using the column density inputs. There are broadly two areas where a non-local model would use information from the wider cloud. For photoionization reactions, the total column density gives the level of visual extinction and therefore the local UV field. Further, calculating the column density of a specific species to the cloud edge corrects their rates for effects like self-shielding. For this time dependent model, the H$_2$ self-shielding and C photoionization rate can simply be calculated using the column density of C and H$_2$ at the start of the time step. Interestingly, despite the fact UCLCHEM treats CO self-shielding, the CO column density was not included as an input. In initial testing, it was found that calculating the CO column density by simply using the CO abundance of the current position and the input column density allowed the model to pass the benchmark tests over a wide range of parameters and visual extinctions.\par
\begin{table}[]
    \centering
    \begin{tabular}{lcc}
    \toprule
        \textbf{Parameter} & \textbf{Symbol} & \textbf{Range}\\
        \midrule
        Gas density & $n_H$ & 10 --\SI{e6}{\per\centi\metre\cubed}\\
        Gas temperature & $T_g$ & 10 -- \SI{e4}{\kelvin} \\
        Dust temperature & $T_d$ & 10 -- \SI{e3}{\kelvin} \\
        Radation Field & $F_{UV}$ & \num{e-1} -- \num{e5} Habing\\
        Cosmic Ray Ionization Rate & $\zeta$ & \num{e-2} -- \num{e5}$\zeta_0$\\
        Total Column density & $N$ & \num{e15} -- \SI{e24}{\per\centi\metre\squared} \\
        H$_2$ Column density & $N_{H2}$ & \num{e5} -- \SI{5e23}{\per\centi\metre\squared} \\
        C Column density & $N_{C}$ & \num{1.0} -- \num{1.5e20} \\
        Metallicity & $Z$  & \num{e-2} -- \num{e0.5}\\
        \bottomrule
    \end{tabular}
    \caption{A list of model inputs excluding the species abundances. $\zeta_0$ indicates a standard cosmic ray ionization rate of  \SI{1.3e-17}{\per\second}. Note that throughout the text the UV field is discussed in units of the Draine field to assist comparison with other codes. However, Chemulator uses Habing (1 Draine = 1.7 Habing).}
    \label{table:inputs}
\end{table}
The other  non-local effect is the calculation of the line opacities of all coolant species. For this, the on-the-spot approximation \citep{Dyson1997TheMedium} was used in which the line opacities are calculated for a homogeneous cloud with the gas properties of the current position and a size that gives the same total column density as the input value. Whilst this does not \emph{a priori} guarantee correct cooling rates, the model is benchmarked extensively in the next section.\par
Overall, this approach means the user needs to provide only three variables to allow the model to capture non-local effects on the temperature and chemistry. An alternative approach would be to make UCLCHEM use local variables only and shift the computation of the local UV field, the rates of photo-processes and the cooling rates to the user. It is possible this would improve the emulator by essentially pre-encoding the relationship between various inputs but would represent a much larger computational burden to the user.\par
The chemical network contains 33 gas phase species, interacting through 330 reactions. The gas phase network is taken from the UMIST12 database \citep{McElroy2013}, including reactions between the network species, cosmic rays and UV photons. X-rays are not considered by this model. In principle, using an emulator could allow complex chemistry to be solved without computational overhead but in an early prototype of this emulator, no working emulator could be produced from a model using a network of 215 species (Appendix~\ref{sec:large-network}). Moreover, the gas temperature for any given set of input parameters was largely unaffected by the choice of network and so, for simplicity, this smaller network was used (see Section ~\ref{sec:consider_this}).\par
The temperature is solved by including it in the ODE system solved for the chemistry. The rate of change of temperature is calculated as 
\begin{equation}
    \frac{dT}{dt} = (\gamma-1) \frac{\Gamma - \Lambda}{k_Bn_H}
\end{equation}
where $\gamma$ is the adiabatic constant of the gas calculated from the number densities of the most abundant gas phase species using Equation 8 from \citet{Grassi2014}. $\Gamma$ and $\Lambda$ are the total heating and cooling rates in \si{\erg\per\centi\metre\cubed\per\second} respectively and $n_H$ is the total gas number density, approximated as the total number density of hydrogen nuclei.
\subsection{Benchmarking the modified UCLCHEM code}
\label{sec:model-benchmarks}
Given the simplified nature of the radiation treatment in this model, the modified UCLCHEM code was benchmarked against UCL\_PDR \citep{Bell2005,Priestley2017ModellingNebula}. UCL\_PDR is a 1D PDR code that uses ray tracing to consistently solve the line opacities for all positions in the model together and to calculate column densities for the treatment of photoionization reactions. Therefore, it is an ideal test of the 0D approximation of the modified UCLCHEM. Moreover, it was itself benchmarked against many other PDR codes \citep{Rollig2007}. For this benchmarking, all cooling rates other than molecular line cooling were also turned off so that both codes shared heating and cooling mechanisms. As a result, any differences can be attributed to either the single point treatment or time dependent chemistry of UCLCHEM.\par
\begin{figure*}
\centering
\includegraphics[width=0.95\textwidth]{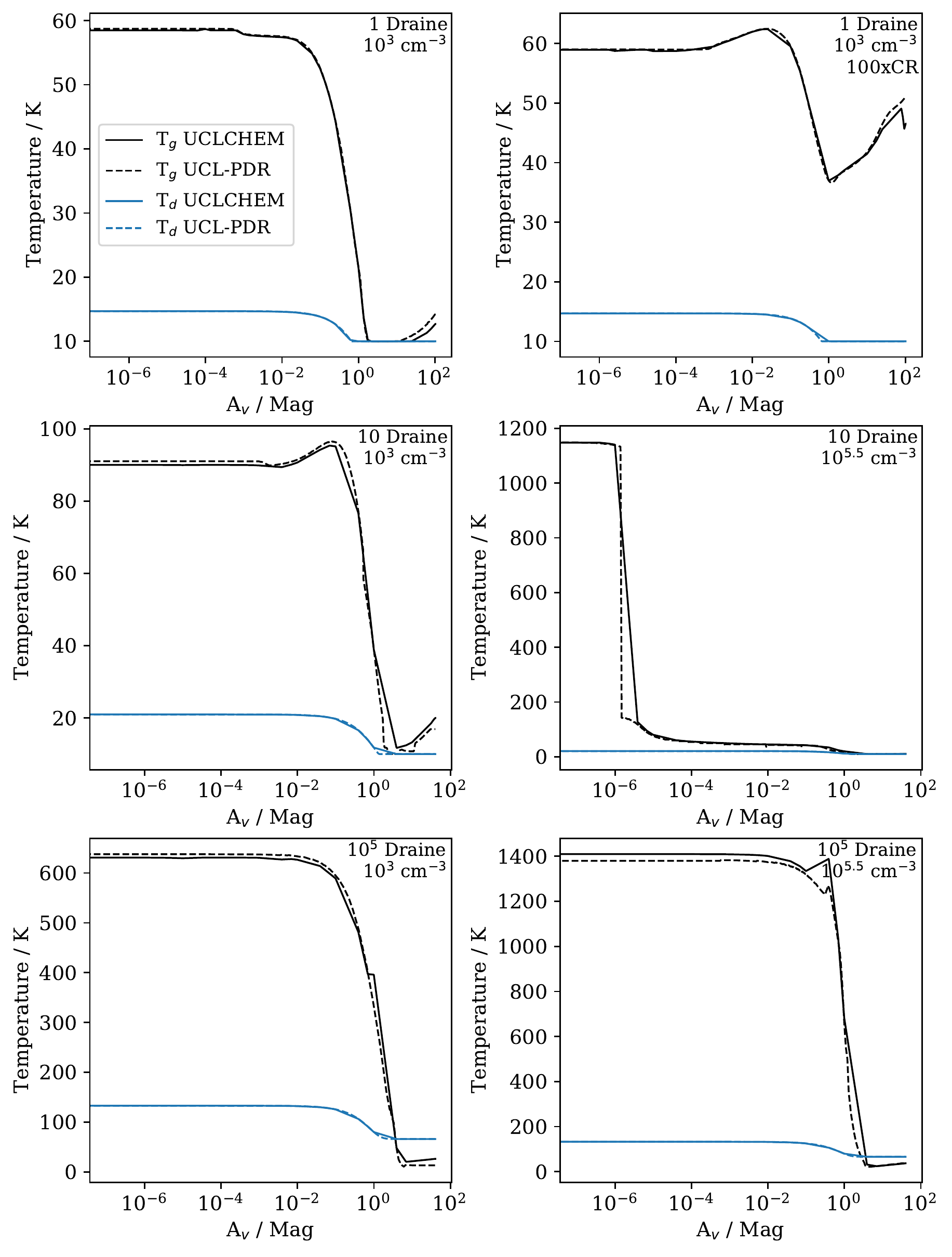}
\caption{Temperature as a function of visual extinction, the lower four plots show \citet{Rollig2007} benchmarking models. In each subplot, the UV field in Draine is noted and in the top right, 100xCR indicates the cosmic ray ionization rate is 100 times standard. The models agree extremely well, allowing for the finer A$_V$ sampling of UCL\_PDR.}
\label{fig:temperature-benchmarks}
\end{figure*}

The two codes were compared using the four benchmark models of \citet{Rollig2007} as well as models with a low radiation field, low metallicity and high cosmic ray ionization rate. Figure~\ref{fig:temperature-benchmarks} shows the results of the \citep{Rollig2007} benchmarks, the low radiation field and high cosmic ray ionization rate models. In every model, the temperatures between the two codes were in close agreement.\par
\begin{figure}
\centering
\includegraphics[width=0.45\textwidth]{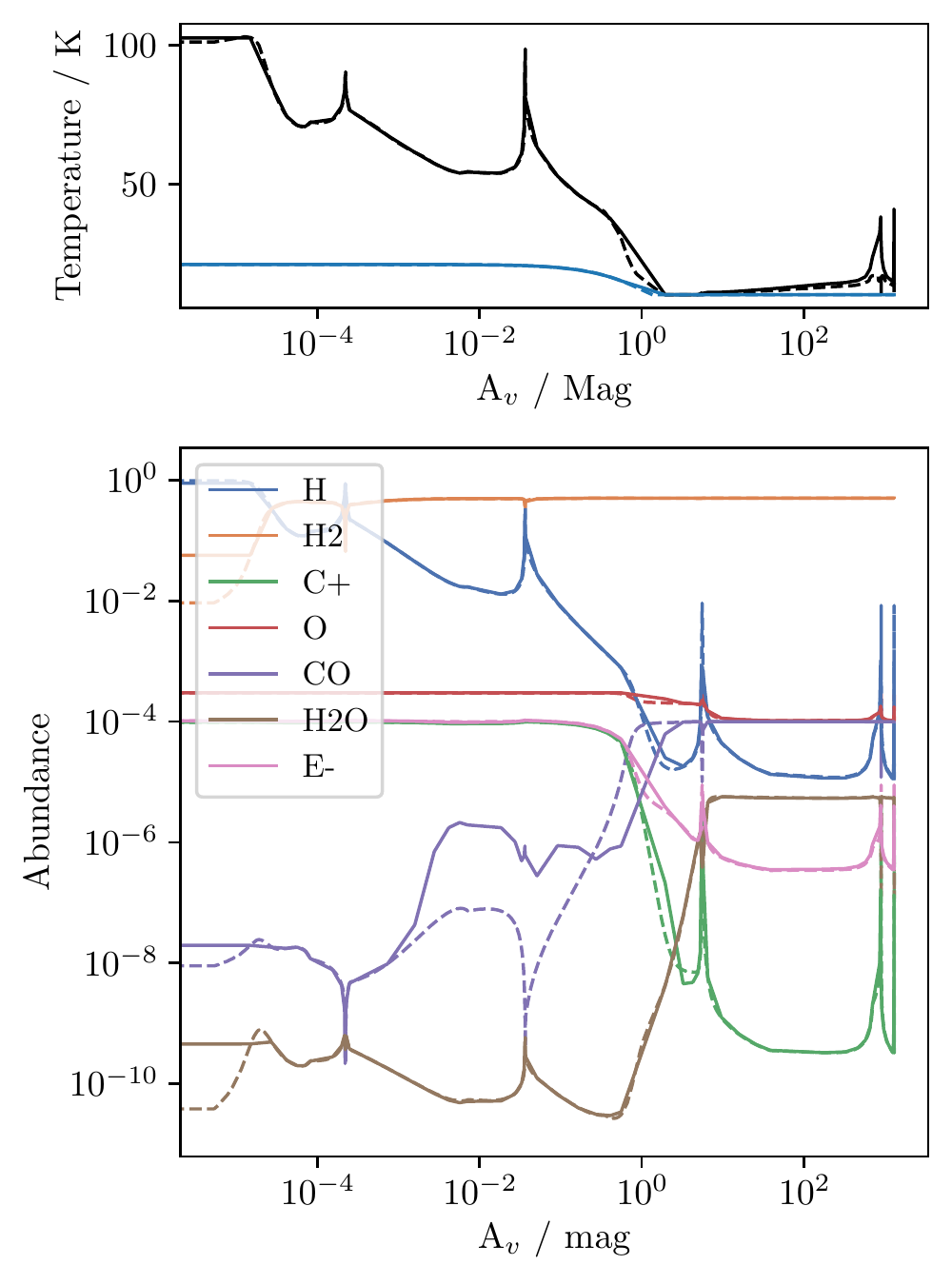}
\caption{Upper: Gas (black) and dust (blue) temperature against $A_v$. Lower: fractional abundance against $A_v$. In both plots, results from UCLCHEM are plotted as solid lines and equivalent values from UCL\_PDR are plotted with dashed lines in the same colour.}
\label{fig:sinewave}
\end{figure}

Each of the \cite{Rollig2007} benchmarks modelled a cloud of constant density. In order to investigate whether this homogeneity was important for UCLCHEM to achieve accurate results in a 0D model, further models were run for clouds with more complex density profiles. In particular, Figure~\ref{fig:sinewave} shows the temperature and abundances of several species as a function of $A_v$ for a cloud where the density at any position is a sine wave function of the distance to the cloud edge. This unphysical scenario would reveal any problems that UCLCHEM suffers when dealing with inhomogeneous clouds. However, it is clear from the figure that the agreement between the two codes is very good.\par 
\subsection{Considerations for the Full Model}
\label{sec:consider_this}
A major conclusion of \citet{Rollig2007} was that a large degree of fine tuning is required to make PDR codes agree. Therefore, it is important to consider the effect of including additional cooling mechanisms and more complex chemistry in the model after the benchmarking tests were completed.\par
In particular, it should be noted that the initial elemental abundances have a large effect on the gas temperature. In the worst case, in the high density benchmark model with a radiation field of \SI{e5}{\draine}, the gas temperature using solar abundances \citep{Asplund2009} reaches \SI{7750}{\kelvin} rather than the \SI{1380}{\kelvin} obtained using the \citet{Rollig2007} benchmarking abundances.\par
However, holding the abundances constant, the benchmarks were performed including the additional cooling mechanisms in Table~\ref{table:cooling} and a larger network of $\sim$ 250 species. This gave a maximum temperature difference of 5\% compared to the benchmarking models. This justifies the use of a smaller network for the emulator.\par
\begin{figure}
\centering
\includegraphics[width=0.45\textwidth]{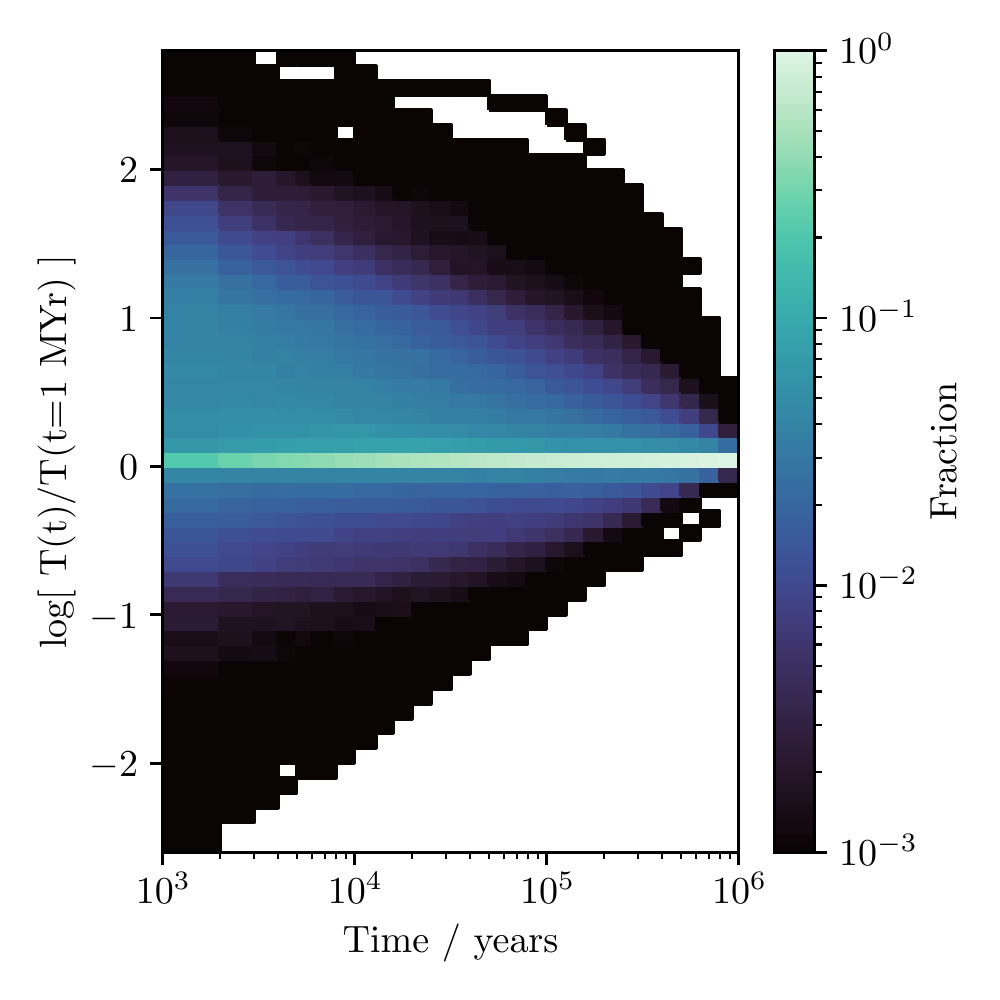}
\caption{A histogram where the colour shows the fraction of models at a given time which have a given log ratio of current temperature to equilibrium temperature. After as few as 1000 years, the majority of models have reached their equilibrium temperature. However, even as the time approaches 1 Myr, $\sim 1\%$ of models have not reached equilibrium which demonstrates that assuming equilibrium will often lead to incorrect chemical abundances.}
\label{fig:equilibrium}
\end{figure}

Finally, it is worth noting the benefits of the single point model over the 1D PDR codes that are available. UCL\_PDR, like many PDR codes, assumes the chemistry and temperature reach equilibrium. In Figure~\ref{fig:equilibrium}, the gas temperature as a fraction of the equilibrium temperature is given for different times for a sample of 20000 random UCLCHEM models (see Section~\ref{sec:data} for the models used). Equilibrium is often reached quickly with 80\% of models reaching equilibrium by 0.1 Myr. However, 5\% of models do not reach equilibrium after 0.5 Myr. Thus for a sub-grid model with short time steps, the model presented here offers a significant improvement over equilibrium models for many cases. Further, given that the heating and cooling processes have been shown to be implemented with adequate accuracy by the benchmarks, the differences in temperature obtained with the full network must in fact be the result of improved chemistry. 
\section{Dimensionality Reduction}
\label{sec:dimensionality}
The inputs for the full model are the 33 initial chemical abundances and the physical variables listed in Table~\ref{table:inputs}. Considering all inputs, there are 41 variables which would need to be sufficiently well sampled to provide adequate training data for the emulator which poses a considerable challenge. Further, an emulator that could take all 41 variables and return them at a time $\Delta t$ later would be required and a regressor with so many outputs is unlikely to be uniformly accurate. \par
\begin{figure}
    \centering
    \includegraphics{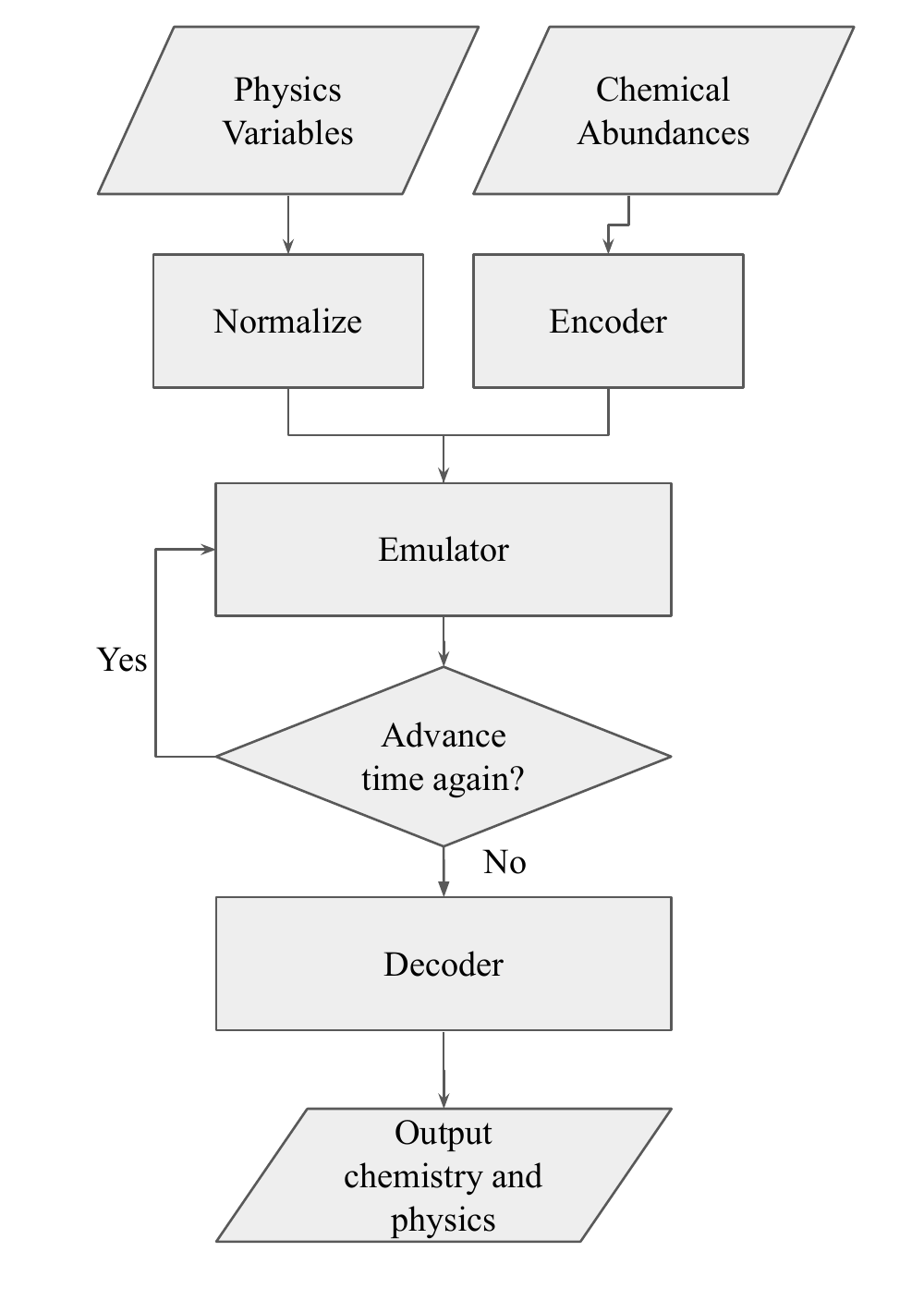}
    \caption{Flowchart showing the operations performed by Chemulator. In particular, it shows how the autoencoder described in Section~\ref{sec:dimensionality} combines with the emulated version of the chemical model to produce the final output.}
    \label{fig:flowchart}
\end{figure}
However, the chemical abundance variables cannot be independent because the chemical network is a closed system of ODEs. Intuitively, the more atoms there are, the fewer molecules there must be and the more ions in the gas phase, the more free electrons there must be. Therefore, it should be possible to transform the abundances to a new, lower dimensional set of variables with minimal loss of accuracy.\par
Using such a dimensionality reduction procedure, the operation of Chemulator would be to first encode the input chemical abundances into the lower dimensional space, then emulate the advancement of the (encoded) chemistry and temperature, before decoding the abundances. This is illustrated in Figure~\ref{fig:flowchart} and the following section details the creation of the encoder and decoder.
\subsection{Dimensionality Dataset}
\label{sec:data}
To do this, a dataset of abundances that gave a good representation of all possible abundances in the model was created. Latin hyper cube sampling (LHS) \citep{McKay1979ACode} was used to efficiently sample the physical parameters in Table~\ref{table:inputs} in log-space within the given ranges to produce initial conditions for 10000 models. An exception to this sampling procedure had to be made for the column densities. Whilst the total column density was sampled with the other parameters, the H$_2$ and C column densities are dependent on the total column density and so could not be sampled independently. To obtain sensible column densities for those molecules, the benchmarking models were examined to find possible ranges for any given total column density. These possible ranges were then randomly sampled to get a H$_2$ and C column density compatible with the total column density for each model.\par
Those models were run for 1 Myr starting with elemental gas and abundances were written out every 1000 yr. Following that, another 10000 samples from the physical parameter space were taken and models were run for 1 Myr starting from the final abundances of a model from the first set. This introduced different chemical histories to the dataset as well as varying the physical parameters. The resulting dataset contained \SI{2e7}{} sets of abundances from various stages of chemical evolution in physical conditions that span the range of those covered by the emulator.\par
This dataset was then used to find a transformation that could transform the abundances to a lower dimensional set of variables and then recover them with minimal losses. Ultimately, an autoencoder was chosen for this purpose. An autoencoder is a neural network which returns the input as an output. By creating an autoencoder where one of the hidden layers is very small, the abundances can be compressed.\par
To train the network, the dataset was transformed in the following way:
\begin{itemize}
    \item All abundances less than \SI{e-20}{} were clipped to \SI{e-20}{}.
    \item The log of the abundance was taken.
    \item Log abundances were scaled to be between 0 and 1.
    \item If any set of scaled log abundances was identical to another to at least 3 decimal places, one set was removed to ensure there were no duplicate sets of abundances in the data.
\end{itemize}
The log of the abundances was used so that no preference was given to the high abundance species. Without this, a 10\% error on a species like H$_2$ would contribute $\sim$ \num{e4} times as much error to the network as even an abundant species like CO.\par
The abundances were clipped to \num{e-20} as this greatly improved the autoencoder accuracy over UCLCHEM's internal clipping at \num{e-30}. This negligible error of at most \num{e-20} on the species abundance was considered an acceptable given that observed abundances in the ISM are typically greater than \num{e-12}.\par
The removal of duplicates improved the efficiency of training the neural network as approximately 40\% of the training data was removed with no loss of information. It also improves the network by preventing it from favouring accurate recovery of a small set of commonly repeated abundances. Finally, scaling variables is a common approach to training neural networks. It also has the advantage that the rms error becomes intuitive as it is the average fractional error on the outputs.\par
\begin{table}[]
    \centering
    \begin{tabular}{ccc}
    \toprule
    Layer & Nodes & Type  \\
    \midrule
    0     & 33 & Input \\
    1     & 256 & Dense - Swish \\
    2     & 8 & Dense - Swish \\
    3     & 256 & Dense - Swish \\
    4     & 33 & Dense - Sigmoid \\
    \bottomrule
    \end{tabular}
    \caption{Description of autoencoder layers and configuration. Layers 0-2 comprise the encoder and layers 2-4 are the decoder.}
    \label{table:autoencoder}
\end{table}
In addition to this pre-processing, two aspects of UCLCHEM could be used to improve the emulator. Firstly, the e$^-$ abundance is always the sum of all ion abundances. Thus, this abundance does not need to be encoded and can be simply recovered from the ions. Secondly, the total H in the model must equal one, this is enforced in the emulator by removing H$^+$ when the total is larger than one and then H if the abundance is still too large. Other species have total abundances that can vary based on metallicty and so no conservation is applied to them.
\subsection{Selecting an Autoencoder}
\label{sec:choose-encoder}
The abundance dataset was split 70:30 into training and validation data and a large grid of neural network configurations with different sizes and numbers of layers and different activation functions was run. Each autoencoder required approximately one minute per epoch to train and 30-50 epochs to reach a minimum validation error. The neural networks were trained by minimizing the mean squared error (MSE) on the recovered abundances. This is the mean squared difference between the input abundances and the predicted abundances. MSE values between \SIrange{e-6}{e-5}{} could be obtained with many neural networks with encoded sizes of as few as 4 variables.\par
However, neural networks can show strongly non-linear behaviour particularly when multiple layers are combined. This would mean small changes to the encoded values could result in large changes to output abundances. Since the emulator will work with these encoded variables, simple behaviour was preferred because the emulator would introduce small errors to the encoded values which should ideally correspond to small errors in the abundances. Therefore, whilst the best autoencoders had two layers in the encoding and decoding parts of the network, the final network chosen was one with only a single layer in the encoder and decoder. The model described in Table~\ref{table:autoencoder} has an MSE of \num{8e-6} on the test data, which corresponds to an RMS error on the predited log abundances of 0.3\%.\par
\begin{figure}
    \centering
    \includegraphics[width=0.5\textwidth]{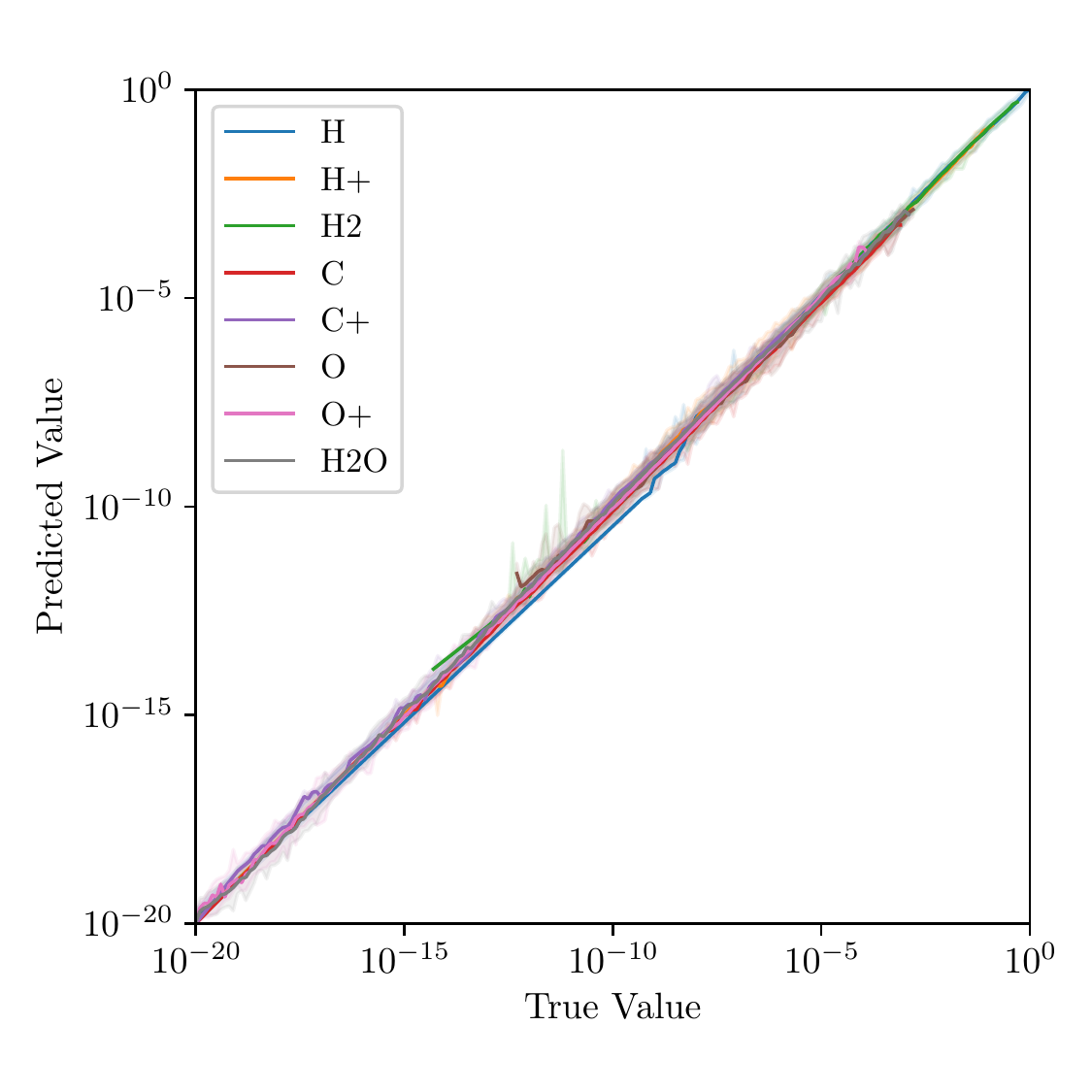}
    \caption{Predicted abundance after a pass through the autoencoder as a function of true abundance for several important species. The solid lines show the median predicted value in the test set for a given real value, the shaded area shows the predicted range for 95\% of encoded abundance sets.}
    \label{fig:autoencoder}
\end{figure}

This accuracy is demonstrated visually in Figure~\ref{fig:autoencoder}. Most species show an extremely tight relationship between the autoencoded and original abundances. The autoencoder's performance is worst when the abundance of a species is much lower than is typical. This can be seen in the spikes in the H$_2$ predictions at $\sim$ \num{e-10}. The figure also shows that at low abundances (\textless \num{e-5}), the H abundance is systematically underpredicted, likely as a result of the enforcement of H conservation described in Section~\ref{sec:data}. Nevertheless, the performance of this autoencoder was considered to be sufficient and therefore it was used for the emulator.\par
\subsection{Drawbacks of the Autoencoder Approach}
\label{sec:drawbacks}
Despite using the simplest network with an acceptable accuracy, the drawback of using an autoencoder was that, upon experimentation, it was found that the chemical abundances could not be sampled in encoded space.  The complexity of this dimensionality reduction method is such that any set of abundances can be accurately encoded and decoded but not all possible combinations of encoded variables will produce realistic or even reasonable abundances.\par
When testing this, the entire dimensionality reduction data was encoded and the range of values each encoded variable took were used as limits on those variable. LHS sampling between those limits was then performed to generate a well sampled set of encoded abundances. However, upon decoding, it was found that many abundance sets were infeasible such as total carbon abundances ten times that of hydrogen meaning that this sampling procedure could not be used to generate well sampled abundances.\par
Given that one of the two motivators for the dimensionality reduction was improved sampling, a simpler dimensionality reduction method was tried. A PCA analysis of the training data was conducted to investigate whether a simple linear transformation of the abundances would suffice. However, the PCA analysis required 12 variables to describe 99\% of the variance in the abundances and this gave an rms error on the log abundances of \num{1.4e-2}. Encoding the chemistry in PCA components required too many components to be useful in order to achieve an acceptable accuracy.\par
In light of this, it was not possible to evenly sample physical and chemical parameter space. Therefore, the dimensionality dataset was used to train the emulator. The chemical and physical parameters for each of the 10000 models in that dataset were written to file every 1000 years from 0 to 1 Myr inclusive. This means there were 1000 pairs of initial parameters with the corresponding values 1000 years later in each model. Thus 20 millions pairs of inputs and outputs were available to train the emulator. \par
Whilst the autoencoder was not used for sampling, it was still used to reduce the input parameters for the emulator. It is likely that an improved dimensionality reduction would represent a major improvement to this work and may even allow a much larger chemical network to be emulated. This will be the focus of later study.

\section{Creating an Emulator}
\label{sec:emulator}
The model presented so far is a complex and accurate time dependent model for the thermodynamics and chemistry of a parcel of gas. However, unlike models such as PRIZMO \citep{Grassi2020} which have been optimized for direct use in hydrodynamic models, it would be inefficient to run as a sub-grid model. The main aim of this work is to produce an emulator for the model that is similarly accurate but has a far lower computational cost than any equivalent model.
\subsection{Training the Emulator}
\label{sec:training}
As noted in Section~\ref{sec:drawbacks}, the dimensionality dataset was used to train the emulator. In fact, the possiblity that this would be a secondary use of the dimensionality dataset is the reason a fixed timestep was used in the generation of that data. By matching each timestep to the one that follows, an emulator can be trained that always advances the temperature and chemistry by 1000 years. Larger timesteps can be obtained by repeatedly running the emulator and an emulator working with small, fixed timesteps is easier to develop than one which can produce variable time advancements.\par
All chemical abundances were encoded by the autoencoder to give eight chemical variables. The physical parameters were then log scaled and then all inputs were min-max scaled to take the range 0-1. To avoid bias, the data were rounded to n decimal places and duplicate rows were removed before returning to the original values. Rounding to two or three decimal places was found to greatly improve the model performance over higher values of n. Finally, this dataset was split 70:30 to train and test the emulator.\par
The emulator is a neural network which takes the 16 input variables and returns the gas temperature and the eight encoded abundance variables. The number of hidden layers, the size and the activation function for each layer was chosen by training a large grid of models, minizing the MSE on the outputs. \par
\begin{figure*}
    \centering
    \includegraphics[width=\textwidth]{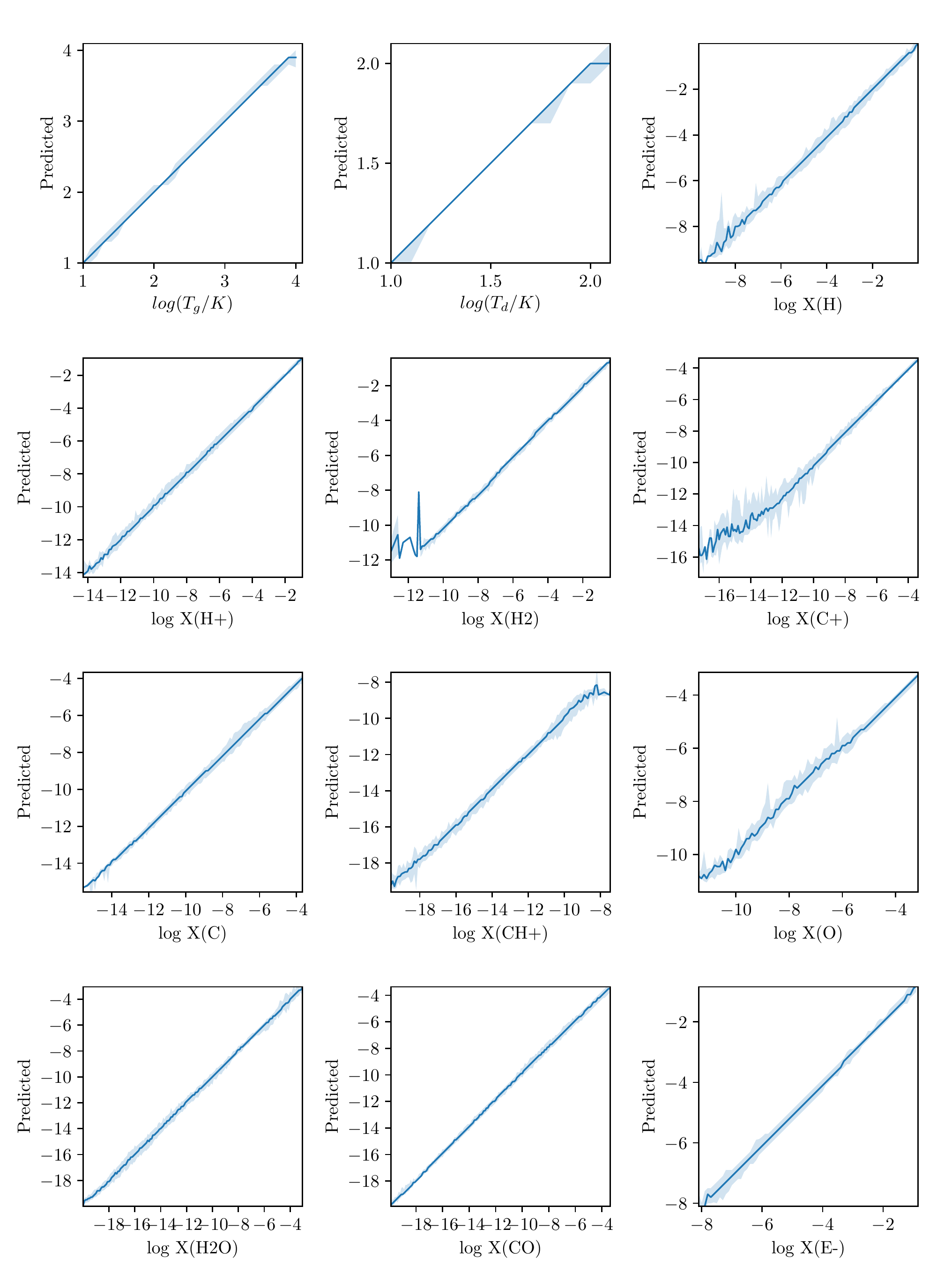}
    \caption{Predicted against real model outputs for the emulator test set. The solid line shows the median value predicted for a given true value and the shaded region shows the range of values predicted in 95\% of cases.}
    \label{fig:emulator_single_runs}
\end{figure*}

It was found that MSE values of \SIrange{e-5}{e-4}{} could be obtained with networks with 2-4 hidden layers of at least 128 nodes each using ReLu or Swish activation functions \citep{Ramachandran2018SearchingFunctions}. The performance of a typical emulator on a single time step is shown in Figure~\ref{fig:emulator_single_runs}. Note that despite the fact it is an analytical function of the emulator inputs, the dust temperature has been included as a target for the emulator for ease of implementation. \par
A selection of species abundances are also shown in Figure~\ref{fig:emulator_single_runs}. Since neither the emulator nor autoencoder are trained to conserve mass, oscillations seen in these abundances do not affect the abundancs of other species. The exception is e$^-$ which has an abundance equal to the total ionization fraction of the gas in UCLCHEM and therefore is calculated by summing the total ion abundances in the emulator. By capturing this abundance well, the emulator is recovering that fraction.\par
However, a known flaw of these kinds of emulators is that over many time steps, error accumulates and the predicted values after n iterations can be extremely far from the true value. This has been seen in similar models in the past \citep[][private comm.]{Grassi2011}. Since this emulator is intended to be used over many iterations of a hydrodynamical model, this would be unacceptable and it was found that all single emulators suffered from this issue. 
\subsection{Error Mitigation}
\label{sec:errormit}
Once it was found that even a small single time step error produced unacceptable error growth, two methods were explored to make the emulator robust to errors that would be introduced by the neural network.\par
Firstly, a Gaussian noise layer was added to the network during both the autoencoder and emulator training. This is a neural network layer which is only active during training and introduces Gaussian noise with some standard deviation $\sigma$. This noise was added to the encoded variables in the autoencoder and to the input of the emulator. If a network can still accurately predict the outputs with this noise, it should be robust to errors of the same order as $\sigma$. Given the typical RMSE of the emulators is 0.01, values of sigma between 0.01 and 0.05 were trialled.
Secondly, an ensemble model was employed. This is simply a model which contains N neural networks each trained individually on the data. They each predict the output and then the mean prediction from all N networks is the model output. In this way, if one network predicts an incorrect value, the others may counter it.\par
\begin{figure*}
    \centering
    \includegraphics[width=\textwidth]{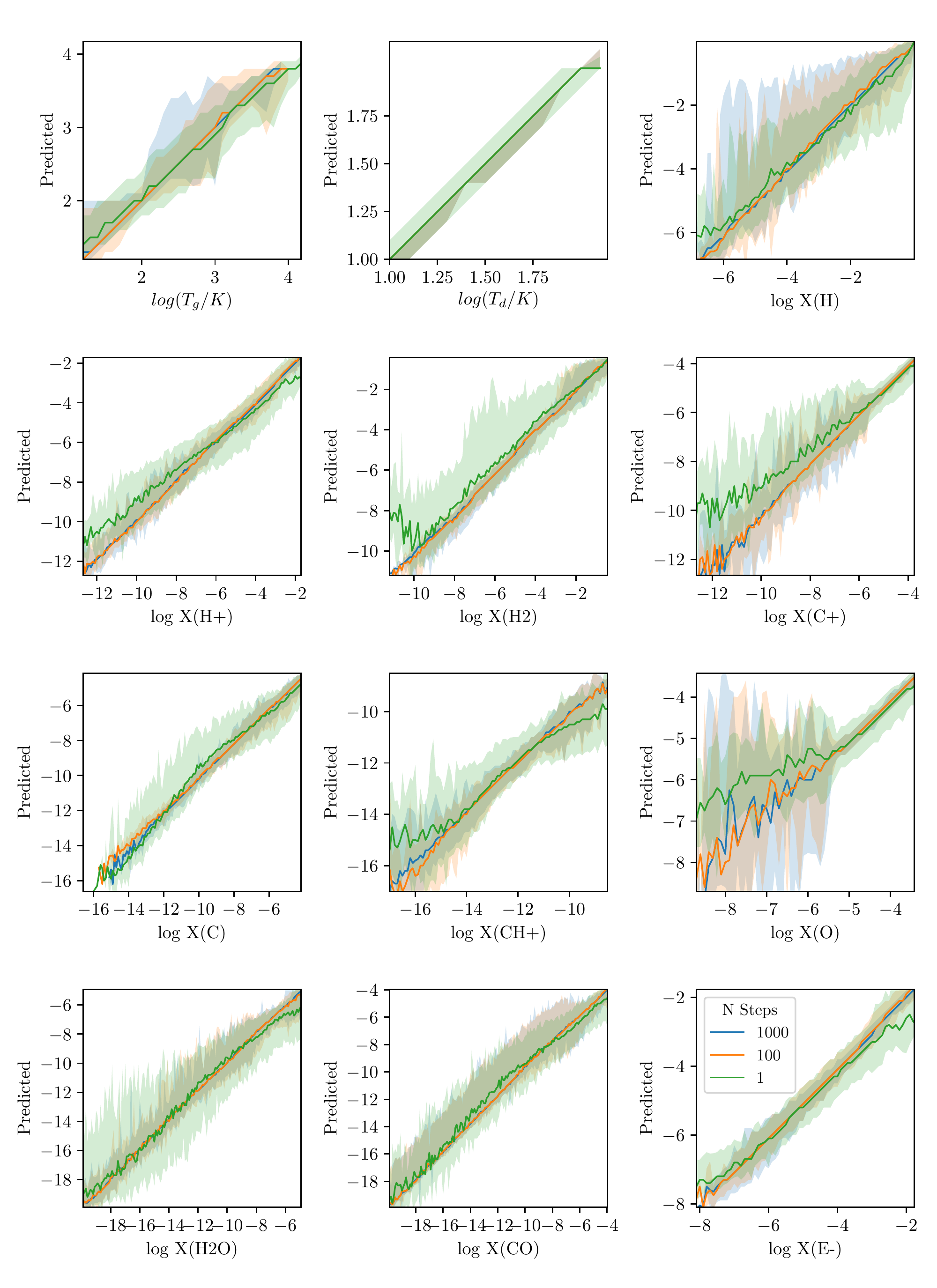}
    \caption{Predicted against real values for time steps in the test set. The median is plotted as a line and the shaded region shows the range predicted by 95\% of models. The colours represent the number of timesteps emulated. For example, the blue lines show predicted evolution of the chemistry for 1 Myr compared to actual value after 1 Myr. Note, the dust temperature error is stable as it is a function of UV and column density which do not change. }
    \label{fig:emulator_many_runs}
\end{figure*}
Increasing the noise was found to improve the model stability at the cost of increasing the single timestep error. Models trained using a $\sigma$ of 0.05 were found to be most stable, giving a similar MSE after 1000 time steps as they give for a single time step. However, those trained with lower noise achieved better MSE error overall and often had a lower MSE error after 1000 time steps than the high noise models despite the fact the error grows with time. It was also found that the ensemble models improved the prediction accuracy but only for ensemble sizes up to four, beyond which the MSE was unchanged.\par
The final ensemble model used four networks each with two ReLu layers of 256 nodes per layer which were trained seperately before being combined using an averaging layer. It was trained with a noise $\sigma$ of 0.02 to minimize error at the cost of some stability as it still performed better after 1000 time steps than a similar, more stable, network trained with a noise of 0.05. The ensemble performs just as well on a single time step as the single networks, with an overall MSE of \num{5.6e-5}. This is equivalent to an RMS error of 0.7\% on the output variables. Most importantly, the error does not continuously grow over many iterations as it did in the noise-free single models.\par
\subsection{Model Performance \& Benchmarking}
The goal of this work is to produce a prototype of a "fast" and "accurate" emulator for thermochemical models. The former criterion is certainly met. The training data required 3750 CPU hours to produce. However, taking the values at $t$=0 yr and emulating each of the 20,000 models to their final time of $t$=\num{e6} years required 5 minutes on a single NVIDIA GeForce GTX 1650 GPU.\par
To demonstrate how well Chemulator meets the second criterion, the performance of the emulator on the test data and a number of simple benchmarks are presented in this section. In Figure~\ref{fig:emulator_many_runs} a similar plot to Figure~\ref{fig:emulator_single_runs} is shown except the predicted and real values are compared after 1, 100, and 1000 time steps to show the error growth. From the size of the shaded regions in the figure, it is clear the error does not grow continuously. One should note, however, that even after 1 time step, the error is larger than in Figure~\ref{fig:emulator_single_runs}. The model appears to perform worst on the first time step of each model in the database, possibly because the random conditions are not always consistent with the abundances. Quantitatively, the MSE on the log-scaled temperature is \num{6e-3} after 1000 yr and \num{6e-3} after 1 Myr.\par
\begin{figure*}
    \centering
    \includegraphics[width=\textwidth]{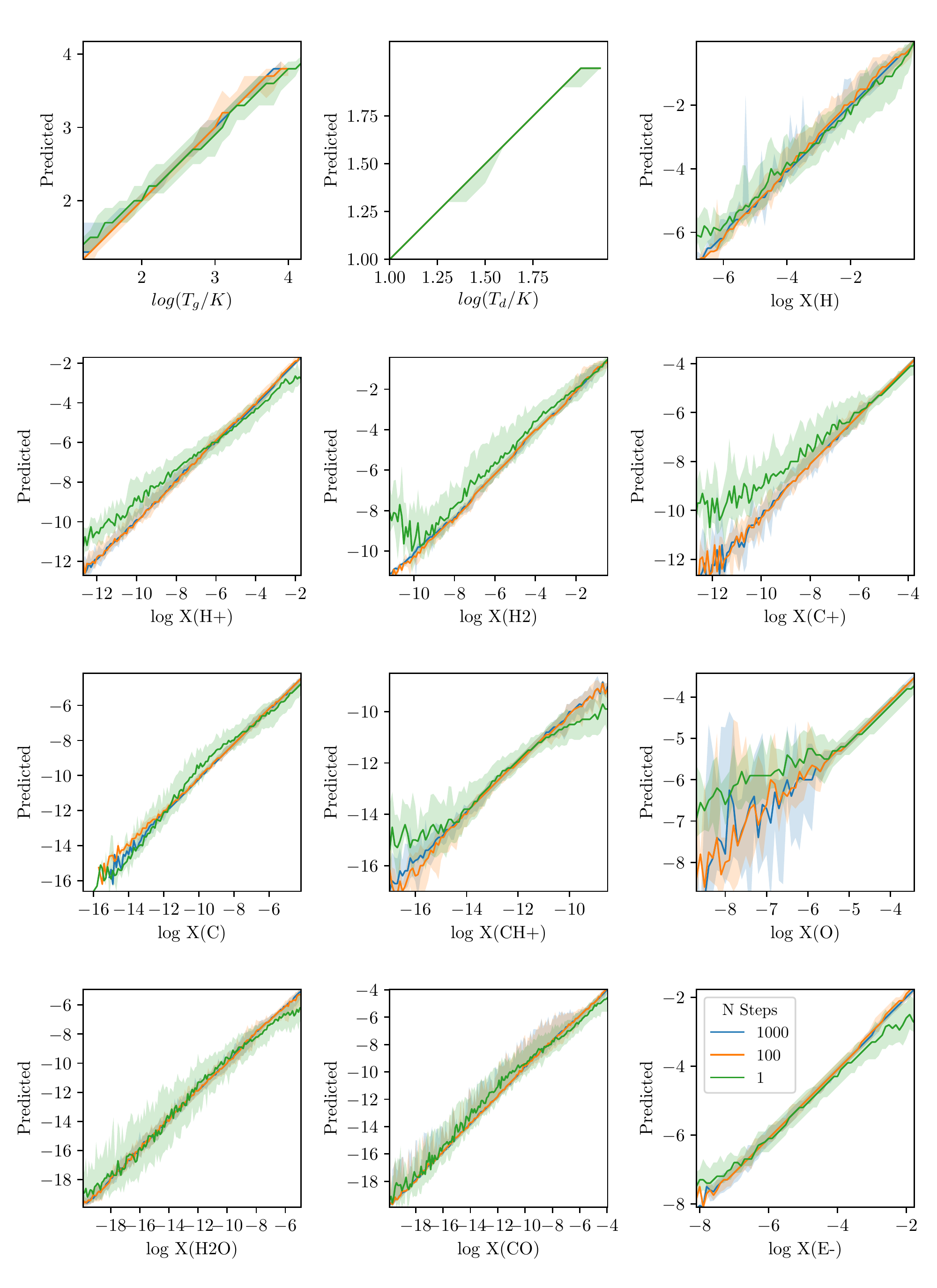}
    \caption{Similar to Figure~\ref{fig:emulator_many_runs} but the shaded region only covers the 67\% of models closest to the correct value rather than 95\%.}
    \label{fig:emulator_many_runs_67}
\end{figure*}
\begin{figure}
    \centering
    \includegraphics[width=0.5\textwidth]{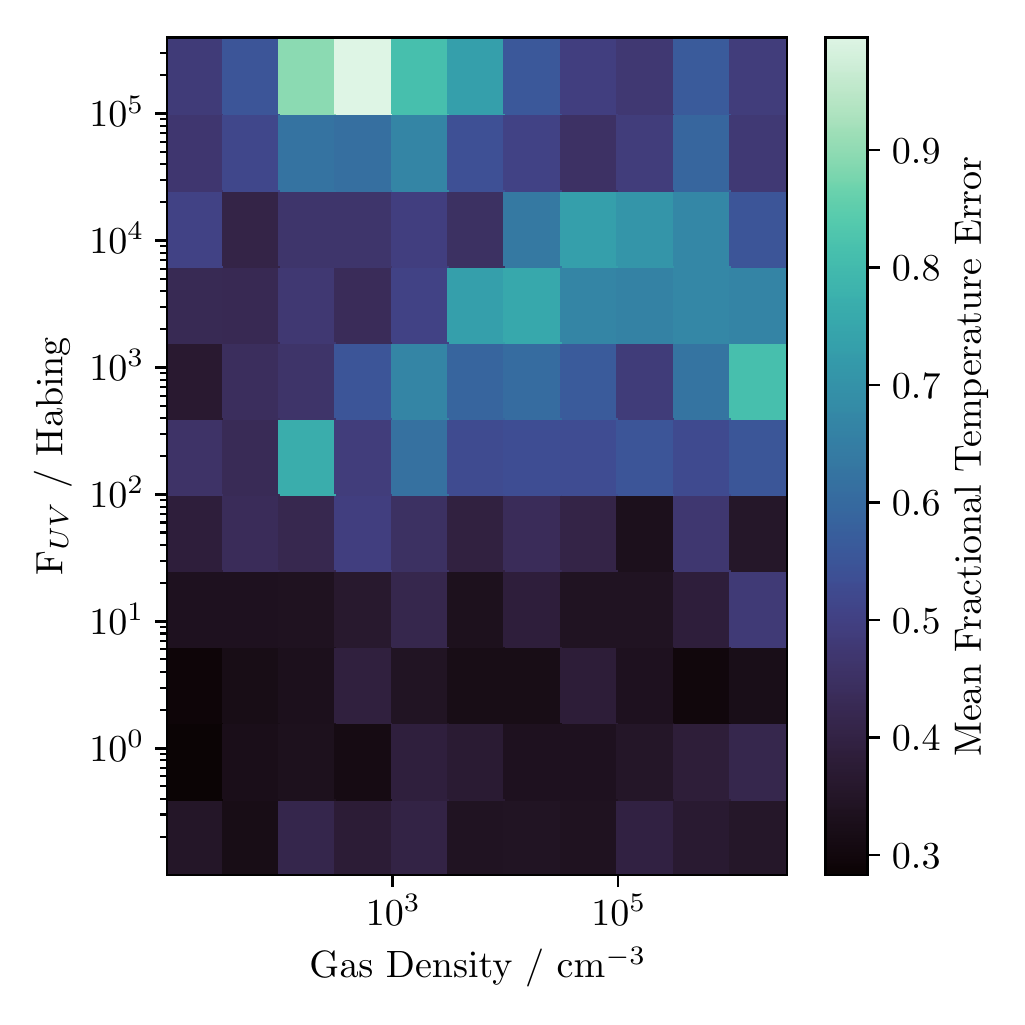}
    \caption{Fractional error on the gas temperature after 1000 iterations (1 Myr) through the emulator as a function of gas density and external UV field.}
    \label{fig:error_dist}
\end{figure}

However, this error is not uniform and a minority of models drastically affect the performance. Figure~\ref{fig:emulator_many_runs_67} is identical to Figure~\ref{fig:emulator_many_runs} except it shows only the 67\% of models closest to the true value. The error distribution is much narrower, especially on the gas temperature, showing that in the majority of cases the emulator produces very accurate temperatures over long times. In order to determine the region of parameter space in which the emulator fails, the error as a function of input parameters was investigated. Figure~\ref{fig:error_dist} shows how the mean fractional error on the temperature varies as a function of radiation field and density after 1 Myr of evolution. The emulator gives better temperature estimates for lower UV, with a notable cut off at a UV field of 100 Draine, above which the average error can be larger than 50\%.\par
The second set of tests considered for the emulator are the \citet{Rollig2007} benchmark models from Section~\ref{sec:model-benchmarks}. Given that the original model passed this benchmarking, an emulator should too. This offers several advantages as a test. Firstly, it is a test many PDR models have passed and therefore any new PDR model should attempt to pass it. Second, the benchmarks are equilibrium models and so the stability of the emulator will be tested. Finally, every single model in the training set uses elemental abundances from \citet{Asplund2009}, scaled by the metallicity. However, the benchmark models use elemental abundances that are not a single scaling of those elemental abundances and so present an interesting out of sample test.\par
\begin{figure*}
    \centering
    \includegraphics[width=\textwidth]{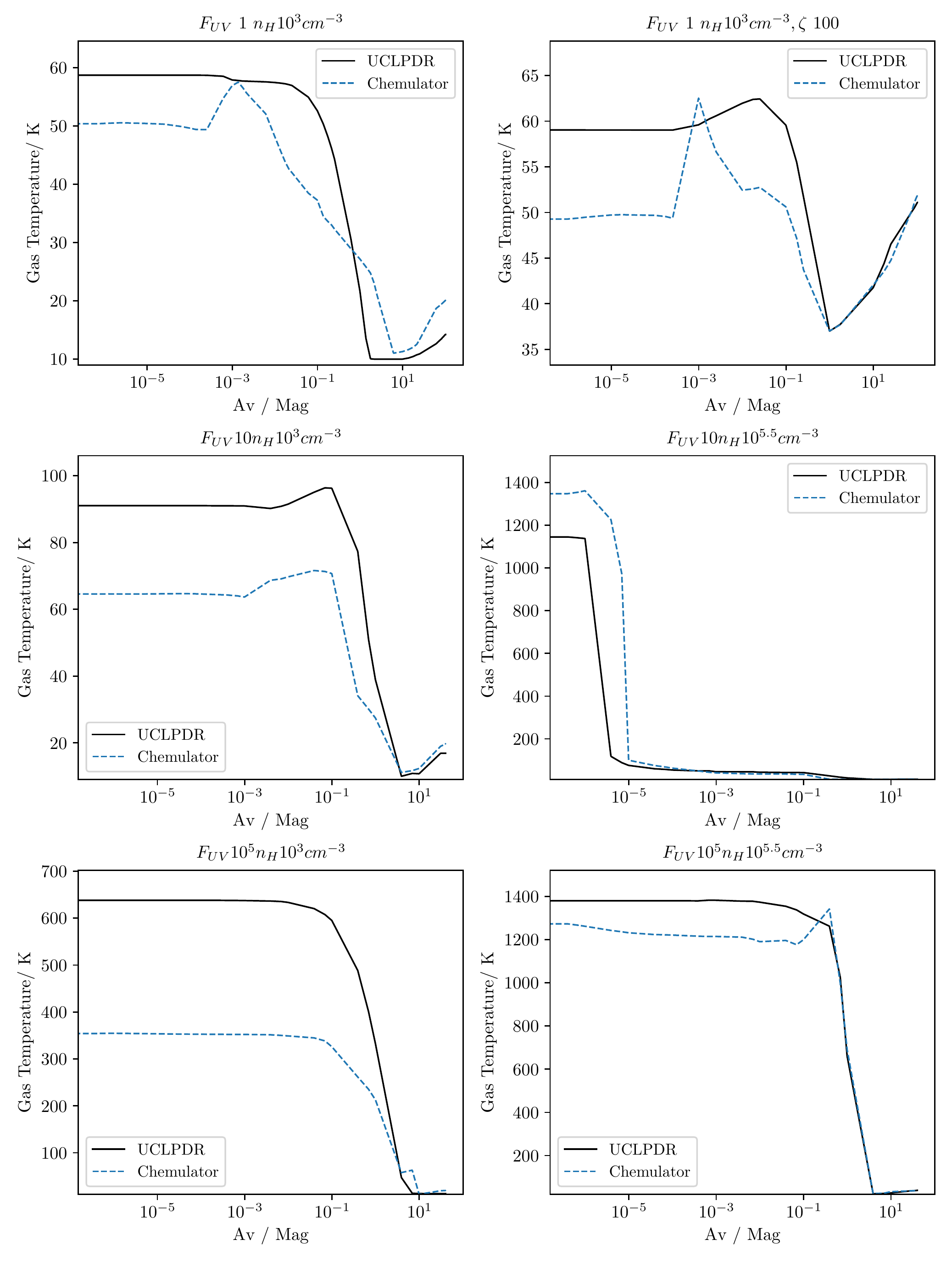}
    \caption{Similar to Figure~\ref{fig:temperature-benchmarks} but comparing gas temperatures from UCL\_PDR and Chemulator. The dust temperature is not included as it is an analytical function of column density and UV.}
    \label{fig:emulator-rollig}
\end{figure*}

Figure~\ref{fig:emulator-rollig} shows the four benchmarking models from \citet{Rollig2007} plus two additional models as in Figure~\ref{fig:temperature-benchmarks}. Comparing the temperature as a function of $A_v$ between UCL\_PDR and Chemulator, it appears the model struggles with low A$_v$ values. At 10-20\% error is typical in these regions and in the low density, high UV model this error is almost 50\% in the low A$_v$ region. This is likely due to the fact the emulator performs worse at higher UV as seen in Figure~\ref{fig:error_dist}.\par 
However, the model does perform well at higher Av. The errors on the temperature are very small once the $A_v \gtrsim$ 1 Mag. Moreover, these solutions are extremely stable and the same result is obtained after 2000 iterations as after the 1000 iterations (1 MYr) shown in these figures.\par
\begin{figure}
\centering
\includegraphics[width=0.45\textwidth]{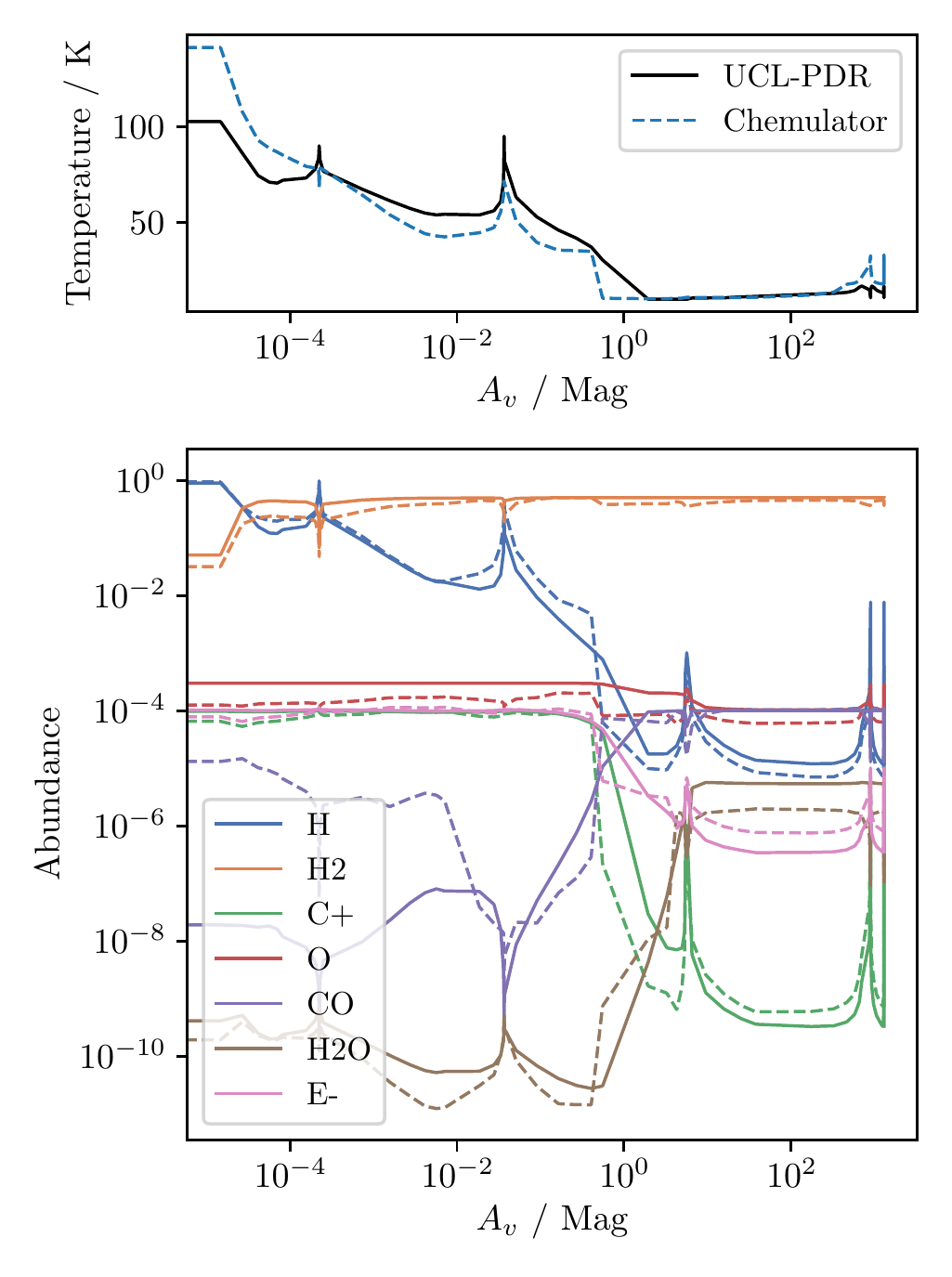}
\caption{Upper: Gas temperature against $A_v$ for a cloud with sinusoidally varying density calculated by UCL\_PDR and emulated by Chemulator. Lower: fractional abundance against $A_v$. In both plots, results from UCL\_PDR are plotted as solid lines and equivalent values Chemulator are dashed.}
\label{fig:emulated-sine}
\end{figure}

This is further demonstrated in Figure~\ref{fig:emulated-sine} which shows the sinusoidally varying cloud used as a complex benchmark for UCLCHEM. Whilst the temperature is over predicted at low A$_v$, Chemulator gives a reasonable estimate further into the cloud. Moreover, the fluctuations in the temperature that arise from the varying density are very well captured.\par
\begin{figure*}
    \centering
    \includegraphics[width=\textwidth]{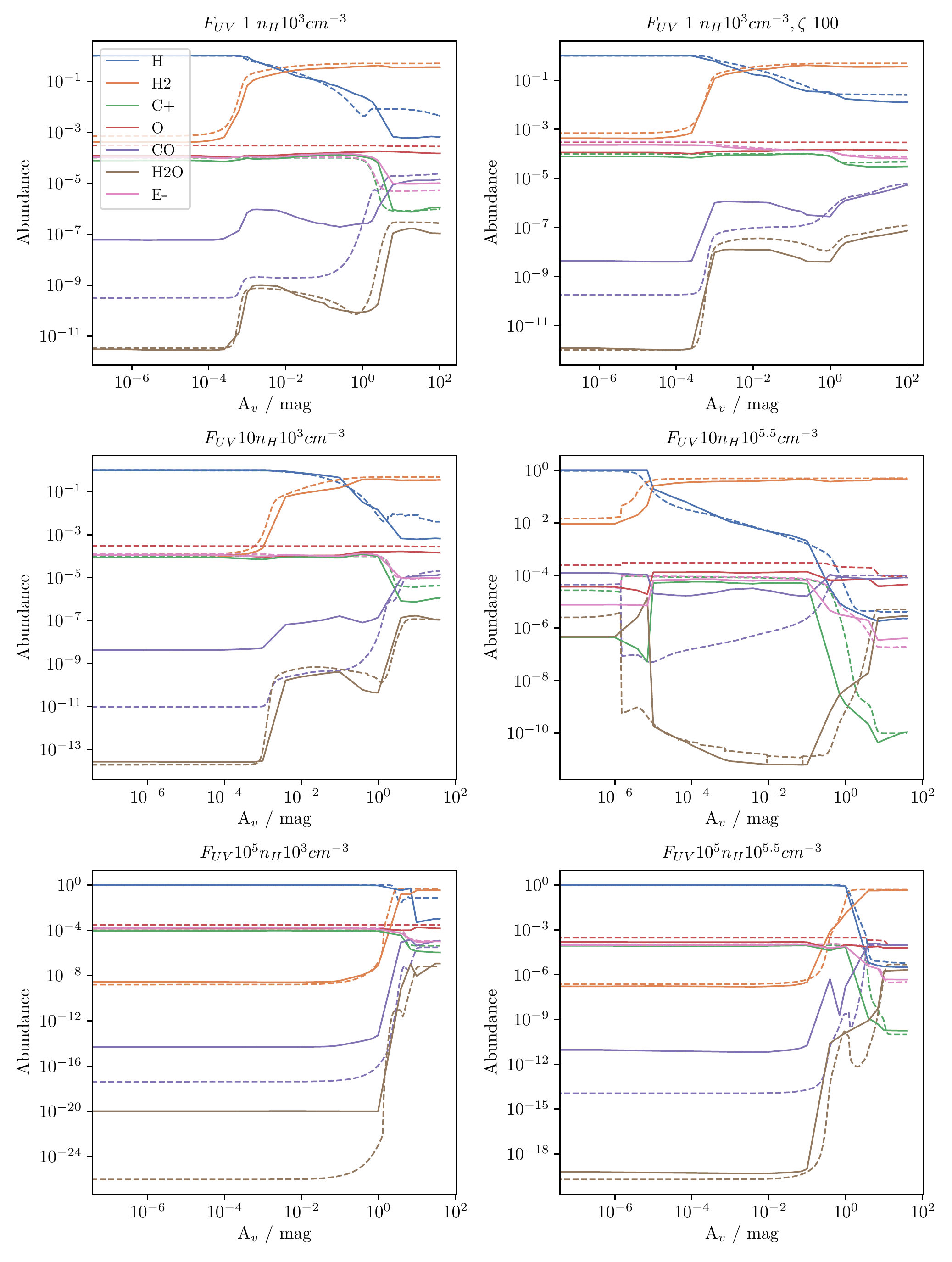}
    \caption{Comparison of the equilibrium abundances obtained with UCL\_PDR and Chemulator for each of the benchmark models. UCL\_PDR abundances are shown as dashed lines and the Chemulator abundances are shown by solid lines.}
    \label{fig:emulator-rollig-chem}
\end{figure*}

Figure~\ref{fig:emulated-sine} also shows the abundances from the sinusoidal model and Figure~\ref{fig:emulator-rollig-chem} presents the abundances from the benchmark models. Chemically, the emulator appears to give accurate estimates of the true abundances. The abundances of most species are close to the benchmark values in every model. Most interestingly, this is even true in the case of the \num{e5} Draine, low density model for which the temperature is poorly captured. It appears Chemulator is able to recover the chemistry even in cases where the temperature predictions are poor.\par
However, Chemulator does overestimate the CO abundance at low A$_v$ where the abundance is low. Since the CO column density is not included in the model inputs, one might expect a poor CO self-shielding treatment is the problem. However, the UCLCHEM benchmarking (eg Figure~\ref{fig:sinewave}) shows UCLCHEM does not suffer from this issue and so it cannot be a result of the inputs. Therefore, this error is more likely due to the fact that CO is most often found at higher abundances and so the emulator struggles with the unusual case of a low CO abundance. Similar behaviour can be seen in Figure~\ref{fig:emulator_many_runs} with species such as O which often have a high abundance. The emulator performs very poorly in situations where a species has a much lower than usual abundance but performs well when the species has a high abundance.\par
\section{Conclusions}
\label{sec:conclusion}
The code Chemulator\footnote{\url{https://github.com/uclchem/Chemulator}} has been presented. Chemulator is an emulator that advances the gas temperature and chemical abundances of a single position in an astrophysical gas. It is very accurate on a single timestep and stable over many iterations with decreased accuracy.\par
An autoencoder was used to reduce the dimensionality of the problem. This was successful in reducing the number of chemical variables in the model from 33 to 8. However, it was found that the encoded space could not be uniformly sampled as it led to spurious abundances. An improved dimensionality reduction procedure could both allow this emulator to be extended to much larger networks and lead to a better sampling of the input parameter space.\par
Chemulator was used to calculate the gas temperature of a standard suite of PDR models. It performed well on these when the visual extinction was high, demonstrating both accuracy and stability. However, it gave large error on at low visual extinctions, often underestimating the temeprature and giving a temperature a factor of two too small in a model with an external UV field of \num{e5} Draine and density of \SI{e3}{\per\centi\metre\cubed}.\par
Given the density and UV field constraints on the performance of Chemulator, it would be a useful code for applications such as large scale ISM modelling. It should be noted that the code to develop these emulators has been released alongside the pre-trained Chemulator. Thus, more specialized applications such as the modelling of planetary atmospheres could be also be served by retraining the emulator for a given parameter space.\par
Overall, Chemulator is a strong first step, demonstrating the promise of this approach. However, improvements need to be made to make it more generally usable as a chemical tool. It is likely that the emulator could be improved by a more accurate or less complex dimensionality reduction. The introduction of a noise layer during training ensured the emulator was resilient to small errors in the encoded variables but it is possible the autoencoder introduces large changes to the encoded variables for small changes in abundance, making it difficult for the emulator to learn the relationships between the encoded variables, the physics, and their subsequent values.\par
Further improvements could be made through the production of a training dataset that is better engineered to cover all realistic inputs. The dimensionality reduction is a part of this but it is also important to investigate sampling techniques which will produce a dataset that uniformly covers the chemical parameter space rather than just the parameter space of the initial physical inputs. It is possible the poor performance at low visual extinctions could be rectified by altering the training set to more strong represent these regions. 
\begin{acknowledgements}
We thank the anonymous referee for their helpful comments which improved this manuscript. This work is part of a project that has received funding from the European Research Council (ERC) under the European Union’s Horizon 2020 research and innovation programme MOPPEX 833460. JDI acknowledges support from the Science and Technology Facilities Council of the United Kingdom (STFC) under ST/T000287/1. The authors thank DiRAC for use of their HPC system which allowed this work to be performed.

\end{acknowledgements}
\appendix
\section{Extending to Larger Networks}
\label{sec:large-network}
An early iteration of this work utilized a network of 215 species interacting through 2508 reactions including both gas and grain surface species. The possibility of producing an emulator which could solve such complex chemistry with a small computation time was an obvious goal. However, no working emulator could be produced and in this appendix, the strengths and failings of that model are discussed.\par
Interestingly, despite the fact this network had almost seven times as many species as the small network, it was possible to produce an autoencoder that had a similar accuracy to the small network encoder without being much larger. A network with one hidden layer of 256 neurons in both the encoder and decoder was used, just like the final autoencoder in section~\ref{sec:choose-encoder}. With an encoded size of 12, rather than 8, chemical variables, the autoencoder for the large network could obtain MSE values $\sim$\num{e-5}. This is promising for future work as it implies even large chemical networks can be represented by very few variables.\par
Following this, a grid of emulators were trained and tested using the encoded chemistry and physical inputs. The best emulator had a single timestep MSE of \num{2.9e-4}, approximately a factor of two larger than the small network emulator. However, even with the introduction of ensemble models and the Gaussian noise layer which ensured the small network emulator had a stable error, error growth could not be prevented in this model.\par
%

%
%
%
\bibpunct{(}{)}{;}{a}{}{,} 
\bibliographystyle{aa}
\bibliography{reduced}

\end{document}